\documentclass{article}
\usepackage[utf8]{inputenc}

\title{khmer: Working with Big Data in Bioinformatics}
\author{Eric McDonald and C. Titus Brown}
\date{February 2013}

\usepackage{url}
\usepackage{graphicx}

\begin{document}

\maketitle

\section{Introduction}

\subsection{Bioinformatics and Big Data}

The field of bioinformatics seeks to provide tools and analyses that
provide understanding of the molecular mechanisms of life on Earth,
largely by analyzing and correlating genomic and proteomic information.
As increasingly large amounts of genomic information, including both
genome sequences and expressed gene sequences, becomes available,
more efficient, sensitive, and specific analyses become critical.

In DNA sequencing, a chemical and mechanical process essentially
``digitizes'' the information present in DNA and RNA. These sequences
are recorded using an \textit{alphabet} of one letter per
nucleotide. Various analyses are performed on this sequence data to
determine how it is structured into larger building blocks and how it
relates to other sequence data. This serves as the basis for the study
of biological evolution and development, genetics, and, increasingly,
medicine.

Data on nucleotide chains comes from the sequencing process in strings of
letters known as \textit{reads}. (The use of the term \textit{read} in the
bioinformatics sense is an unfortunate collision with the use of the term in
the computer science and software engineering sense. This is especially true as
the performance of reading reads can be tuned, as we will discuss. To
disambiguate this unfortunate collision we refer to sequences from genomes as
\textit{genomic reads}.) To analyze larger scale structures and processes,
multiple genomic reads must be fit together. This fitting is different than a
jigsaw puzzle in that the picture is often not known \textit{a priori} and that
the pieces may (and often do) overlap. A further complication is introduced in
that not all genomic reads are of perfect fidelity and may contain a variety of
errors, such as insertions or deletions of letters or substitutions of the
wrong letters for nucleotides. While having redundant reads can help in the
assembly or fitting of the puzzle pieces, it is also a hindrance because of
this imperfect fidelity in all of the existing sequencing technologies. The
appearance of erroneous genomic reads scales with the volume of data and this
complicates assembly of the data.

% TODO: Reference NIH study with info on sequencing cost and Moore's Law.
%	(Suggested by Alexis.)

As sequencing technology has improved, the volume of sequence data being
produced has begun to exceed the capabilities of computer hardware employing
conventional methods for analyzing such data. (Much of the state-of-the-art in
sequencing technology produces vast quantities of genomic reads, typically tens
of millions to billions, each having a sequence of 50 to 100 nucleotides.) This
trend is expected to continue and is part of what is known as the \textit{Big
Data} \cite{web:bigdata} problem in the high performance computing (HPC),
analytics, and information science communities.  With hardware becoming a
limiting factor, increasing attention has turned to ways to mitigate the
problem with software solutions.  In this chapter, we present one such software
solution and how we tuned and scaled it to handle terabytes of data.

Our research focus has been on efficient {\em pre-processing}, in which
various filters and binning approaches trim, discard, and bin the
genomic reads, in order to improve downstream analyses.  This approach
has the benefit of limiting the changes that need to be made to downstream
analyses, which generally consume genomic reads directly.

% TODO: can also reference grant proposals at ged.msu.edu/interests.html.

In this chapter, we present our software solution and describe how we
tuned and scaled it to efficient handle increasingly large amounts of
data.

\subsection{What is the khmer Software?}

``khmer'' is our suite of software tools for pre-processing large
amounts of genomic sequence data prior to analysis with conventional
bioinformatics tools \cite{web:khmer} -- no relation to the ethnic
group indigenous to Southeast Asia.  This name comes by free
association with the term 'k-mer': as part of the pre-processing,
genetic sequences are decomposed into overlapping substrings of a
given length, \textit{k}. As chains of many molecules are often called
\textit{polymers}, chains of a specific number of molecules are called
\textit{k-mers}, each substring representing one such chain.  Note
that, for each genomic read, the number of k-mers will be the number
of nucleotides in the sequence minus $k$. So, nearly
every genomic read will be decomposed into many overlapping k-mers.

\begin{figure}[ht!]
\centering
\includegraphics[scale=0.5]{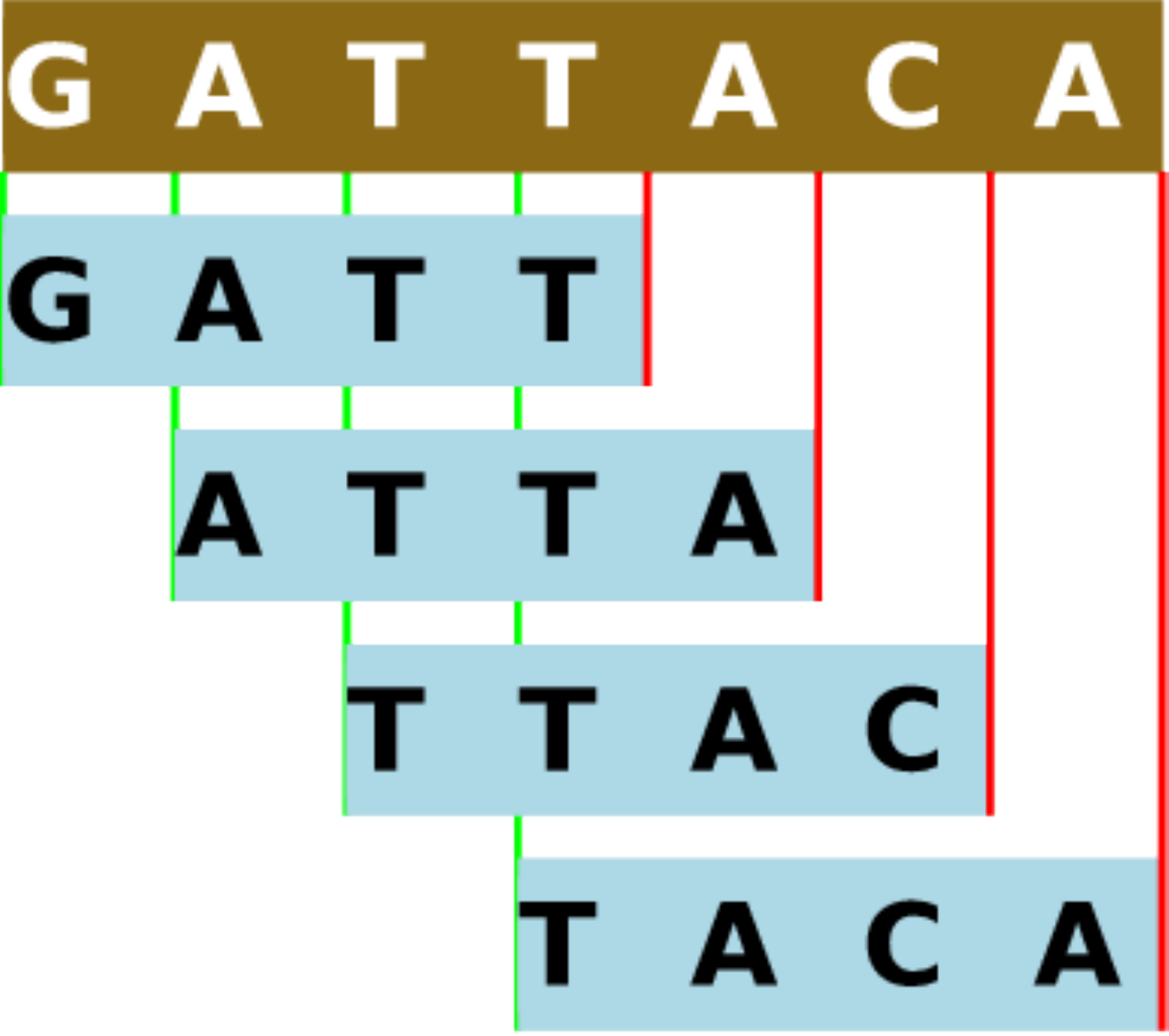}
\caption{Decomposition of Genomic Sequence into 4-mers.  In khmer, the forward sequence and reverse complement of each k-mer are hashed to the same value, in recognition that DNA is double-stranded.  See Future Directions.}
\label{kmers}
\end{figure}

% TODO: Add citations for diginorm and partitioning papers.

\begin{figure}[ht!]
\centering
\includegraphics[scale=0.5]{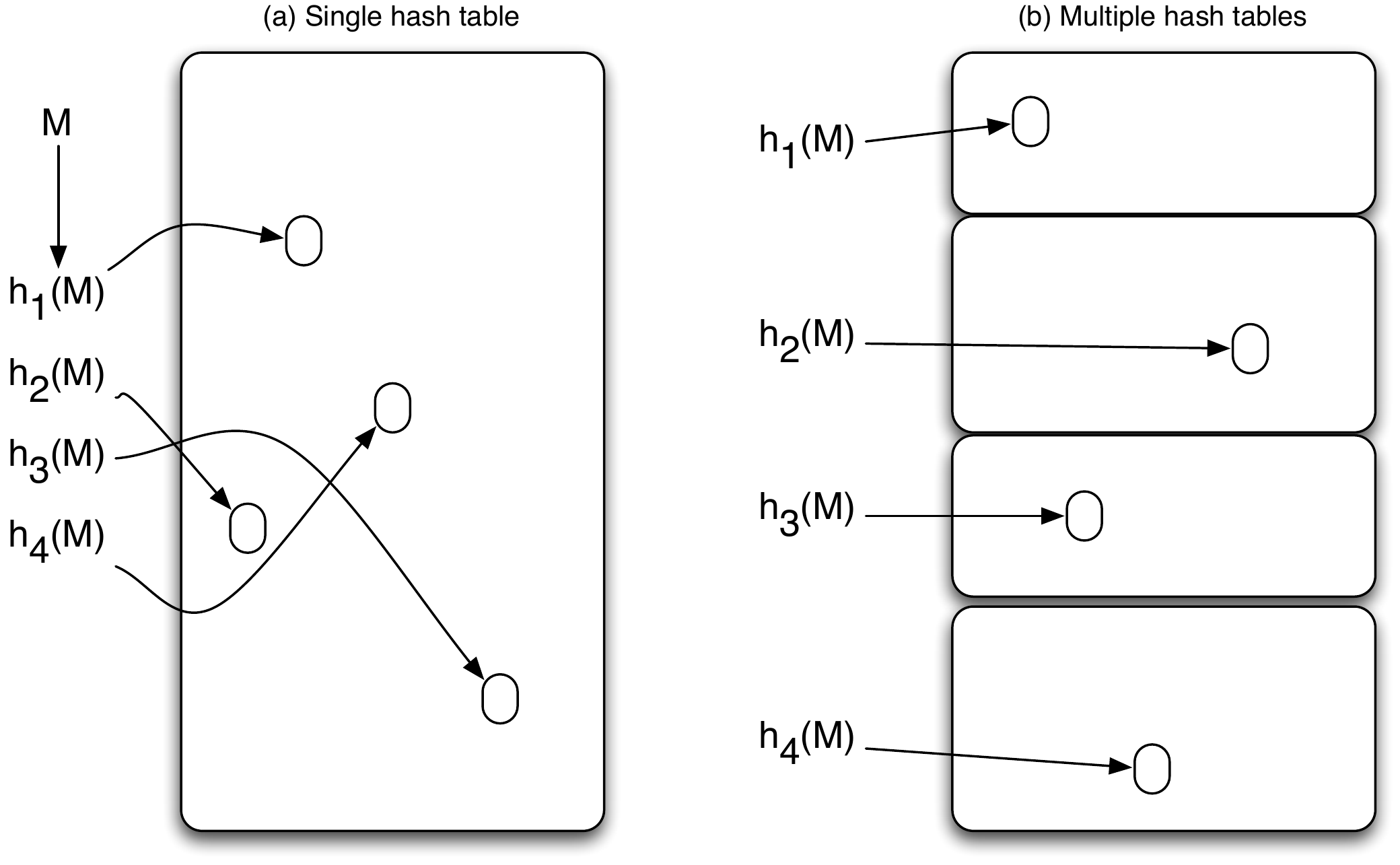}

\caption{A Bloom filter is essentially a large, fixed-size hash
  table, into which elements are inserted or queried using multiple
  orthogonal hash functions, with no attempt at collision tracking;
  they are therefore {\em probabilistic} data structures.  Our
  implementation uses multiple distinct hash tables each with its own
  hash function, but the properties are identical.  We typically
  recommend that khmer's Bloom filters be configured to use as much
  main memory as is available, as this reduces collisions maximally.}

\label{bloomFilter}
\end{figure}

Since we want to tell you about how we measured and tuned this piece of open
source software, we'll skip over much of the theory behind it. Suffice it to
say that k-mer counting is central to much of its operation. To compactly count
a large number of k-mers, a data structure known as a \textit{Bloom filter}
\cite{web:BloomFilter} is used (Figure \ref{bloomFilter}). Armed with k-mer counts, we can then exclude
highly redundant data from further processing, a process known as
``digital normalization''. We can also treat low
abundance sequence data as probable errors and exclude it from further
processing, in an approach to error trimming.  These normalization and
trimming processes greatly reduce
the amount of raw sequence data needed for further analysis, while mostly
preserving information of interest.

khmer is designed to operate on large data sets of millions to billions of
genomic reads, containing tends of billions of unique k-mers.  Some of our
existing data sets require up to a terabyte of system memory simply to
hold the k-mer counts in memory, but this is not due to inefficient
programming: in \cite{kmer-percolation} we show that khmer is considerably more
memory efficient than any exact set membership scheme for a wide regime
of interesting k-mer problems.  It is unlikely that significant improvements
in memory usage can be obtained easily.

Our goal, then, is simple: in the face of these large data sets, we
would like to optimize khmer for processing time, including most especially
the time required to load data from disk and count k-mers.

For the curious, the khmer sources and documentation can be cloned
from GitHub at \url{http://github.com/ged-lab/khmer.git}.  Khmer has
been available for about four years, but only with the posting of
several preprint papers have others started to use it; we estimate the
user population at around 100 groups based on e-mail interactions in
2012, although it seems to be growing rapidly as it becomes clear
that a large class of assembly problems is more readily
tractable with khmer \cite{diginorm}.

\section{Architecture and Performance Considerations}

Khmer started as an exploratory programming exercise and is evolving into
more mature research code over time.
From its inception, the focus has been on solving
particular scientific problems with as much as accuracy or ``correctness'' as
possible. Over time, as the software has come into greater use around the
world, issues such as packaging, performance, and scalability have become more
prominent. These issues were not necessarily neglected in earlier times, but
they now have a higher profile than they once did. Our discussion will center
around how we have analyzed and solved particular performance and scaling
challenges.
Beacuse khmer is research code still under development, it routinely
receiving new features and
has a growing collection of impermanent scripts built up around it.  We must
be careful to ensure that changes made to
improve performance or scalability do not break existing
interfaces or to cause the software not decrease in accuracy or
correctness. For this reason, we have proceeded along a strategy that
combines automated testing with careful,
incremental optimization and parallelization. In conjunction with other
activities pertaining to the software, we expect this process to be essentially perpetual.

\begin{figure}[ht!]
\centering
\includegraphics[scale=0.6]{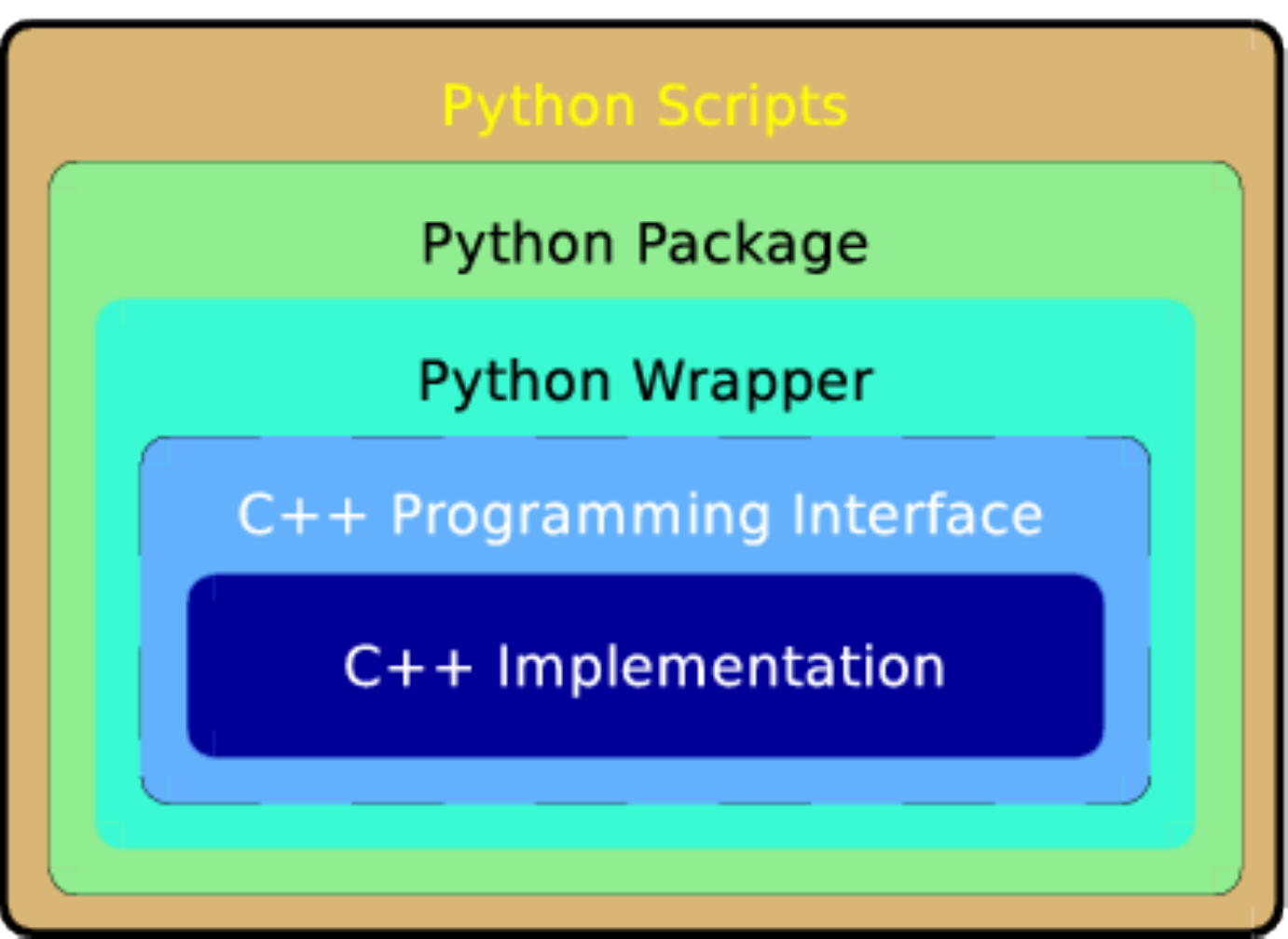}
\caption{A Layered View of the khmer Software}
\label{khmerLayers}
\end{figure}

The core of the software is written in C++. This core consists of a data pump (the
component which moves data from online storage into physical RAM), parsers for
genomic reads in several common formats, and several k-mer counters. An
\textit{application programming interface} (API) is built around the core. This
API can, of course, be used from C++ programs, as we do with some of our test
drivers, but also serves as the foundation for a Python wrapper. A Python
package is built upon the Python wrapper. Numerous Python scripts are
distributed along with the package. Thus, the khmer software, in its totality,
is the combination of core components, written in C++ for speed, higher-level
interfaces, exposed via Python for ease of manipulation, and an assortment of
tool scripts, which provide convenient ways to perform various bioinformatics
tasks.

The khmer software supports batch operation in multiple phases, each with
separate data inputs and outputs. For example, it can take a set of genomic
reads, count k-mers in these, and then, optionally, save the Bloom filter hash
tables for later use. Later, it can use saved hash tables to perform k-mer
abundance filtering on a new set of genomic reads, saving the filtered data.
This flexibility to reuse earlier outputs and to decide what to keep allows a
user to tailor a procedure specific to his/her needs and storage constraints.

\begin{figure}[ht!]
\centering
\includegraphics[scale=0.4]{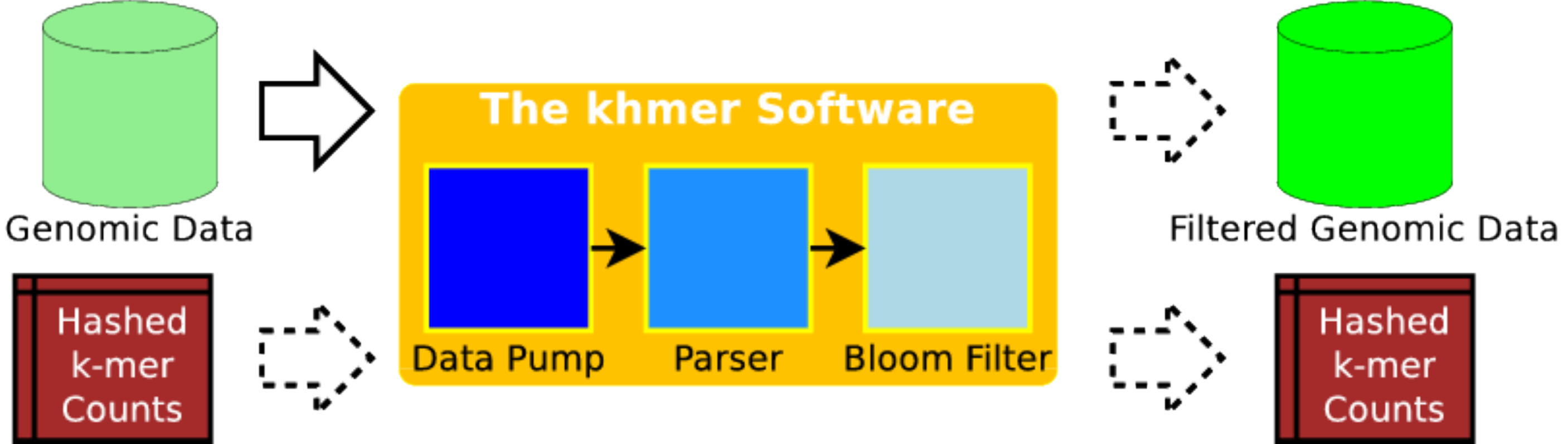}
\caption{Data Flow through the khmer Software}
\label{khmerDataFlow}
\end{figure}

Lots and lots of data (potentially terabytes) must be moved from disk to memory
by the software. Having an efficient data pump is crucial, as the input
throughput from storage to CPU may be 3 or even 4 orders of magnitude less than
the throughput for data transfer from physical RAM to CPU. For some kinds of
data files, a decompressor must be used. In either case, a parser must work
efficiently with the resultant data. The parsing task revolves around
variable-length lines, but also must account for invalid genomic reads and
preserving certain pieces of biological information, which may be exploited
during later assembly, such as pairings of the ends of sequence fragments. Each
genomic read is broken up into a set of overlapping k-mers and each k-mer is
registered with or compared against the Bloom filter.  If a previously stored
Bloom filter is being updated or used for comparison, then it must be loaded
from storage.  If a Bloom filter is being created for later use or updated,
then it must be saved to storage.

The data pump always performs sequential access on files and can potentially be
asked to read large chunks of data at one time. With this in mind, the
following are some of the questions which come to mind:
\begin{itemize}
\item Are we fully exploiting the fact that the data is accessed sequentially?
\item Are enough pages of data being prefetched into memory to minimize access
latency?
\item Can asynchronous input be used instead of synchronous input?
\item Can we efficiently bypass system caches to reduce buffer-to-buffer copies
in memory?
\item Does the data pump expose data to the parser in a manner that
does not create any unnecessary accessor or decision logic overhead?
\end{itemize}

Parser efficiency is essential, as data enters in a fairly liberal
string format and must be converted into an internal representation before
any further processing is done.  Since each individual data record is
relatively small (100-200 bytes), but there are millions to billions of
records, we have focused quite a bit of effort on optimizing the
record parser. The parser, at its core, is a loop which breaks up the 
data stream into genomic reads and stores them in records, performing some 
initial validation along the way.

Some considerations regarding parser efficiency are:
\begin{itemize}
\item Have we minimized the number of times that the parser is touching the 
data in memory?
\item Have we minimized the number of buffer-to-buffer copies while parsing 
genomic reads from the data stream?
\item Have we minimized function call overhead inside the parsing loop?
\item The parser must deal with messy data, including ambiguous bases,
too-short genomic reads, and character case.  Is this DNA sequence validation being done as efficiently as possible?
\end{itemize}

For iterating over the k-mers in a genomic read and hashing them, we could ask:
\begin{itemize}
\item Can the k-mer iteration mechanism be optimized for both memory and speed?
\item Can the Bloom filter hash functions be optimized in any way?
\item Have we minimized the number of times that the hasher is touching the 
data in memory?
\item Can we increment hash counts in batches to exploit a warm cache?
\end{itemize}

\section{Profiling and Measurement}

Simply reading the source code with an eye on performance revealed a number of
areas for improvement. However, we wanted to systematically quantify the amount
of time spent in various sections of the code. To do this, we used several
profilers: the GNU Profiler (gprof) and the Tuning and Analysis Utilities
(TAU). We also created instruments within the source code itself, allowing a
fine granularity view of key performance metrics.

\subsection{Code Review}

Blindly applying tools to measure a system (software or otherwise) is rarely a
good idea. Rather, it is generally a good idea to gain some understanding of
the system before measuring it. To this end, we reviewed the code by eye first.

Manually tracing the execution paths of an unfamiliar code is a good idea.
(One of the authors, Eric McDonald, was new to the khmer software at the time
he joined the project and he did this.) While it is true that profilers (and
other tools) can generate call graphs, those graphs are only abstract
summaries. Actually walking the code paths and seeing the function calls is a
much more immersive and enlightening experience. Debuggers can be used for such 
walks, but do not readily lend themselves to the exploration of code paths less 
travelled. Also, moving through an execution path step-by-step can be quite 
tedious. Breakpoints can be used for testing whether certain points in the 
code are hit during normal execution, but setting them requires some 
\textit{a priori} knowledge of the code. As an alternative, the use of an 
editor with multiple panes works quite well. Four display panes can often 
simultaneously capture all of the information a person needs to know - and is 
mentally capable of handling - at any given point.

The code review showed a number of things, some, but not all, of which were
later corroborated by profiling tools. Some of the things we noticed were:

\begin{itemize}
\item We expected the highest traffic to be in the k-mer counting logic.
\item Redundant calls to the \texttt{toupper} function were present in the highest traffic 
regions of the code.
\item Input of genomic reads was performed line-by-line and on demand and without 
any readahead tuning.
\item A copy-by-value of the genomic read struct performed for every parsed and valid genomic read.
\end{itemize}

Although the preceding may seem like fairly strong self-criticism, we
would like to stress that a greater emphasis had been placed on
utility and correctness of khmer up to this point.  Our goal was to optimize
existing and mostly correct software, not to redevelop it from scratch.

\subsection{Tools}

Profiling tools primarily concern themselves with the amount of time spent in
any particular section of code. To measure this quantity, they inject
instrumentation into the code at compile time. This instrumentation does change
the size of functions, which may affect inlining during optimization.  The
instrumentation also directly introduces some overhead on the total execution
time; in particular, the profiling of high traffic areas of code may result in
a fairly significant overhead. So, if you are also measuring the total elapsed
time of execution for your code, you need to be mindful of how profiling itself
affects this. To gauge this, a simple external data collection mechanism, such
as \texttt{/usr/bin/time}, can be used to compare non-profiling and profiling
execution times for an identical set of optimization flags and operating
parameters.  

We gauged the effect of profiling by measuring the difference between
profiled and non-profiled code across a range of $k$ sizes -- smaller
$k$ values lead to more $k$-mers per genomic read, increasing
profiler- specific effects.  For $k = 20$, we found that non-profiled
code ran about 19\% faster than profiled code, and, for $k = 30$, that
non-profiled code ran about 14\% faster than profiled code.

% @Eric -- do you have specific numbers to put in this next paragraph?

Prior to any performance tuning, our profiling data showed that the k-mer
counting logic was the highest traffic portion of the code, as we had predicted
by eye. What was a little surprising was how significant of a fraction it was,
contrasted to I/O operations against storage. Given that our trial data sets
were about 500 MB and 5 GB, we did not anticipate seeing much in the way of
cache effects.\footnote{If the size of a data cache is larger than the data
being used in I/O performance benchmarks, then retrieval directly from the
cache rather than the original data source may skew the mesurements from
successive runs of the benchmarks. Having a data source larger than the data
cache helps guarantee data cycling in the cache, thereby giving the appearance
of a continuous stream of non-repeating data.} Indeed, when we controlled for
cache effects, we found that they did not amount to more than a couple of
seconds at most and were thus not much larger than the error bars on our total
execution times.  This left us with the realization that I/O was not our
primary bottleneck at that juncture in the code optimization process.

Once we began parallelizing the khmer software, we wrote some driver programs,
which used OpenMP \cite{web:OpenMP}, to test our parallelization of various
components. While gprof is good at profiling single-threaded execution, it
lacks the ability to trace per-thread execution when multiple threads are in
use and it does not understand parallelization machinery, such as OpenMP. For
C/C++ codes, OpenMP parallelization is determined by compiler pragmas. GNU
C/C++ compilers, in the version 4.x series, honor these pragmas if supplied
with the \texttt{-fopenmp} switch. When OpenMP pragmas are being honored, the
compilers inject thread-handling instrumentation at the locations of the
pragmas and around the basic blocks or other groupings with which they are
associated.

As gprof could not readily give us the per-thread reporting and the OpenMP
support that we desired, we turned to another tool. This was the Tuning and
Analysis Utilities (TAU) \cite{web:TAU} from a collaboration led by the
University of Oregon. There are a number of parallel profiling tools out there
- many of them focus on programs using MPI (Message Passing Interface)
libraries, which are popular for some kinds of scientific computing tasks. TAU
supports MPI profiling as well, but as MPI is not really an option for the
khmer software in its current manifestation, we ignored this aspect of TAU.
Likewise, TAU is not the only tool available for per-thread profiling. The
combination of per-thread profiling and the ability to integrate closely with
OpenMP is one of the reasons that it was appealing to us. TAU is also entirely
open source and not tied to any one vendor.

Whereas gprof relies solely upon instrumentation injected into source code at
compile time (with some additional bits linked in), TAU provides this and other
instrumentation options as well. These options are library interposition
(primarily used for MPI profiling) and dynamic instrumentation of binaries. To
support these other options, TAU provides an execution wrapper, called
\texttt{tau\_exec}. Compile-time instrumentation of source code is supported
via a wrapper script, called \texttt{tau\_cxx.sh}.

TAU needs additional configuration to support some profiling activities.
To get tight OpenMP integration, for example, TAU needs to be
configured and built with support for OPARI. Similarly, to use the
performance counters exposed by newer Linux kernels, it needs to be
configured and built with support for PAPI. Also, once TAU is built,
you will likely want to integrate it into your build system for
convenience. For example, we setup our build system to allow the
\texttt{tau\_cxx.sh} wrapper script to be used as the C++ compiler
when TAU profiling is desired. If you attempt to build and use TAU,
you will definitely want to read the documentation. While much more
powerful than gprof, it is not nearly as facile or intuitive.

\subsection{Manual Instrumentation}

Examining the performance of a piece of software with independent, external
profilers is a quick and convenient way to learn something about the execution
times of various parts of software at a first glance. However, profilers are
generally not so good at reporting how much time code spends in a particular
spinlock within a particular function or what the input rate of your data is.
To augment or complement external profiling capabilities, manual
instrumentation may needed. Also, manual instrumentation can be less intrusive
than automatic instrumentation, since you directly control what gets observed.
To this end, we created an extensible framework to internally measure things
such as throughputs, iteration counts, and timings around atomic or
fine-grained operations within the software itself. As a means of keeping
ourselves honest, we internally collected some numbers that could be compared
with measurements from the external profilers.

For different parts of the code, we needed to have different sets of metrics.
However, all of the different sets of metrics have certain things in common.
One thing is that they are mostly timing data and that you generally want to
accumulate timings over the duration of execution. Another thing is that a
consistent reporting mechanism is desirable. Given these considerations, we
provided an abstract base class, \texttt{IPerformanceMetrics}, for all of our
different sets of metrics. The \texttt{IPerformanceMetrics} class provides some
convenience methods: \texttt{start\_timers}, \texttt{stop\_timers}, and
\texttt{timespec\_diff\_in\_nsecs}. The methods for starting and stopping
timers measure both elapsed real time and elapsed per-thread CPU time. The
third method calculates the difference between two standard C library
\texttt{timespec} objects in nanoseconds, which is of quite sufficient
resolution for our purposes.

To ensure that the overhead of the manually-inserted internal instrumentation
is not present in production code, we carefully wrapped it in conditional
compilation directives so that a build can specify to exclude it.

\section{Tuning}

Making software work more efficiently is quite a gratifying experience,
especially in the face of trillions of bytes passing through it. Our narrative
will now turn to the various measures we took to improve efficiency. We
divide these into two parts: optimization of the reading and parsing of input
data and optimization of the manipulation and writing of the Bloom filter
contents.

\subsection{General Tuning}

Before diving into some of the specifics of how we tuned the khmer software, we
would like to briefly mention some options for general performance tuning.
Production code is often built with a set of safe and simple optimizations
enabled; these optimizations can be generally proven not to change the
semantics of the code (i.e., introduce bugs) and only require a single
compilation pass. Compilers do provide additional optimization options,
however. These additional options can be broadly categorized as
\textit{aggressive optimizations}, which is a fairly standard term in compiler
literature, and \textit{profile-guided optimizations} (PGO) \cite{web:PGO}.
(The two categories are not mutually-exclusive, strictly speaking, but
typically involve different approaches.)

Aggressive optimizations may be unsafe (i.e., introduce bugs) in some cases or
actually decrease performance in other cases. Aggressive optimizations may be
unsafe for a variety of reasons, including sloppiness in floating-point
accuracy or assumptions about different operands being associated with
different memory addresses. Aggressive optimizations may also be specific to a
particular CPU family. Profile-guided optimizations rely on profiling
information to make more educated guesses on how to optimize a program during
compilation and linking.  One frequently-seen profile-guided optimization is
the optimization of locality - attempting to place highly-correlated functions
as neighbors inside the text segment of the executable image so that they will
be loaded into the same memory pages together at runtime.

At this stage in our project, we have avoided both categories of additional
optimizations in favor of targeted algorithmic improvements - improvements that
provide benefits across many different CPU architectures.  Also, from the
standpoint of build system complexity, aggressive optimizations can create
portability issues and profile-guided optimizations add to the total number of
moving parts which may fail. Given that we do not distribute pre-compiled
executables for various architectures and that our target audience is usually
not too savvy about the intricacies of software development or build systems,
it is likely that we will continue avoiding these optimizations until we feel
that the benefits outweigh the drawbacks. In light of these considerations, our
main focus has been on improving the efficiency of our algorithms rather than
other kinds of tuning.

\subsection{Data Pump and Parser Operations}

Our measurements showed that the time spent counting k-mers dominated the time
performing input from storage. Given that interesting fact, it may seem like we
should have devoted all of our efforts to improving the Bloom filter's
performance. But, taking a look at the data pump and parser was worthwhile for
several reasons. One reason was that we needed to alter the design of the
existing data pump and parser to accommodate their use by multiple threads to
achieve scalability. Another reason was that we were interested in reducing
memory-to-memory copies, which could impact the efficiency of the Bloom filter
at its interface with the parser. A third reason is that we wanted to position
ourselves to provide an aggressive readahead or prefetch of data, in case we
were able to improve the efficiency of the k-mer counting logic to the point
that input time became competitive with counting time. Unrelated to performance
tuning, there were also issues with maintainability and extensibility.

As it turns out, all of the above reasons converged on a new design. We will
discuss the thread-safety aspects of this design in more detail later. For now,
we will focus upon the reduction of memory-to-memory copies and the ability to
perform fairly aggressive prefetching of data.

Typically, when a program retrieves data from a block storage device (e.g., a
hard disk), a certain number of the blocks are cached by the operating system,
in case the blocks are needed again. There is some time overhead associated
with this caching activity; also, the amount of data to prefetch into the cache
cannot be finely tuned. Furthermore, the cache cannot be accessed directly by a
user process and so must be copied from the cache into the address space of the
user process.  This is a memory-to-memory copy. 

Some operating systems, such as Linux, allow for their readahead windows to be
tuned some. One can make calls to \texttt{posix\_fadvise(2)} and
\texttt{readahead(2)} for a particular file descriptor, for example. However,
these allow rather limited control and do not bypass caching. We are interested
in bypassing the cache maintained by the OS. This cache actually can be
bypassed if a file is opened with the \texttt{O\_DIRECT} flag and the file
system supports it. Using direct input is not entirely straightforward, as the
reads from storage must be multiples of the storage medium's block size and
must be placed into an area of memory, which has a base address that is a
multiple of the same block size. This requires a program to perform
housekeeping which a file system would normally do. We implemented direct
input, including the necessary housekeeping. There are, however, some cases
where direct input will not work or is otherwise undesirable. For those cases,
we still attempt to tune the readahead window.  Our access of storage is
sequential and we can tell the operating system to read further ahead than it
normally would by using \texttt{posix\_fadvise(2)} to provide a hint.

Minimizing buffer-to-buffer copies is a challenge shared between the data pump
and the parser. In the ideal scenario, we would read once from storage into our
own buffer and then scan our buffer once per genomic read to demarcate a
sequence with an offset and length within the buffer. However, the logic for
managing the buffer is complex enough and the logic for parsing (accounting for
our particular nuances) is complex enough that maintaining an intermediary line
buffer is quite useful for programmer comprehension. To reduce the impact of
this intermediary buffer, we encourage the compiler to rather aggressively
inline this portion of the code. We may ultimately eliminate the intermediary
buffer if performance of this particular region of the code becomes a big
enough issue, but that may come at the expense of an understandable software
design.

\subsection{Bloom Filter Operations}

Recalling that we are working with sequences composed of an alphabet of four
letters: A, C, G, and T, you might ask whether these are uppercase or lowercase
letters. Since our software operates directly on user-provided data, we cannot
rely on the data to be consistently upper- or lower-case, since
both sequencing platforms and other software packages may alter the case.
While this is easy to fix for individual genomic reads, we need to repeat
this for each base in millions or billions of read!

Prior to performance tuning the code was insensitive to case right up
to the points where it validated the DNA string and where it generated the hash
codes. At these points, it would make redundant calls to the C library's
\texttt{toupper} function to normalize the sequences to uppercase, using macros
such as the following:

\begin{verbatim}

#define is_valid_dna(ch) \
    ((toupper(ch)) == 'A' || (toupper(ch)) == 'C' || \
     (toupper(ch)) == 'G' || (toupper(ch)) == 'T')

\end{verbatim}

and:

\begin{verbatim}

#define twobit_repr(ch) \
    ((toupper(ch)) == 'A' ? 0LL : \
     (toupper(ch)) == 'T' ? 1LL : \
     (toupper(ch)) == 'C' ? 2LL : 3LL)

\end{verbatim}

If you read the manual page for the \texttt{toupper} function or inspect the
headers for the GNU C library, you might find that it is actually a
locale-aware function and not simply a macro. So, this means that there is the
overhead of calling a potentially non-trivial function involved - at least when
the GNU C library is being used. But, we are working with an alphabet of four
ASCII characters. A locale-aware function is overkill for our purposes.  So,
not only do we want to eliminate the redundancy but we want to use something
more efficient.

We decided to normalize the sequences to uppercase letters prior to validating
them. (And, of course, validation happens before attempting to convert them
into hash codes.) While it might be ideal to perform this normalization in the
parser, it turns out that sequences can be introduced to the Bloom filter via
other routes. So, for the time being, we chose to normalize the sequences
immediately prior to validating them. This allows us to drop all calls to
\texttt{toupper} in both the sequence validator and in the hash encoders.

Considering that terabytes of genomic data may be passing through the sequence
normalizer, it is in our interests to optimize it as much as we can. One
approach is:

\begin{verbatim}
#define quick_toupper( c ) (0x60 < (c) ? (c) - 0x20 : (c))
\end{verbatim}

For each and every byte, the above should execute one compare, one branch, and
possibly one addition. Can we do better than this? As it turns out, yes. Note
that every lowercase letter has an ASCII code which is 32 (hexadecimal 20)
greater than its uppercase counterpart and that 32 is a power of 2. This means
that the ASCII uppercase and lowercase characters differ by a single bit only.
This observation screams ``bitmask!''

\begin{verbatim}
c &= 0xdf; // quicker toupper
\end{verbatim}

The above has one bitwise operation, no compares, and no branches. Uppercase
letters pass through unmolested; lowercase letters become uppercase. Perfect,
just we wanted. For our trouble, we gained about a 13\% speedup in the
runtime of the entire process (!)

Our Bloom filter's hash tables are... ``expansive''. To increment the counts
for the hash code of a particular k-mer means hitting almost $N$ different
memory pages, where $N$ is the number of hash tables allocated to the filter.
In many cases, the memory pages which need to be updated for the next k-mer are
entirely different than those for the current one. This can lead the much
cycling of memory pages from main memory without being able to utilize the
benefits of caching. If we have a genomic read with a 79-character long
sequence and are scanning k-mers of length 20, and if we have 4 hash tables,
then up to 236 (59 * 4) different memory pages are potentially being touched.
If we are processing 50 million reads, then it is easy to see how costly this
is. What to do about it?

One solution is to batch the hash table updates. By accumulating a number of
hash codes for various k-mers and then periodically using them to increment
counts on a table-by-table basis, we can greatly improve cache utilization.
Initial work on this front looks quite promising and, hopefully, by the time
you are reading this, we will have fully integrated this modification into our
code. Although we did not mention it earlier in our discussion of measurement
and profiling, \texttt{cachegrind}, a program which is part of the open-source
Valgrind \cite{web:Valgrind} distribution, is a very useful tool for gauging
the effectiveness of this kind of work.

\section{Parallelization}

With the proliferation of multi-core architectures in today's world, it is
tempting to try taking advantage of them. However, unlike many other problem
domains, such as computational fluid dynamics or molecular dynamics, our Big
Data problem relies on high throughput processing of data -- it
must become essentially I/O-bound beyond a certain point of parallelization.
Beyond this point, throwing
additional threads at it does not help as the bandwidth to the storage media is
saturated and the threads simply end up with increased blocking or I/O wait
times. That said, utilizing some threads can be useful, particularly if the
data to be processed is held in physical RAM, which generally has a much higher
bandwidth than online storage. As discussed previously, we have implemented a
prefetch buffer in conjunction with direct input. Multiple threads can use this
buffer; more will be said about this below. I/O bandwidth is not the only
finite resource which multiple threads must share. The hash tables used for
k-mer counting are another one. Shared access to these will also be discussed
below.

\subsection{Thread-safety and Threading}

Before proceeding into details, it may be useful to clear up a couple items
about terminology. People often confuse the notion of something being
thread-safe with that of something being threaded. If something is thread-safe,
then it can be simultaneously accessed by multiple threads without fear of
corrupted fetches or stores. If something is multi-threaded, then it is
simultaneously operated by multiple threads of execution.

As part of our parallelization work, we remodeled portions of the C++ core
implementation to be thread-safe without making any assumptions about a
particular threading scheme or library. Therefore, the Python
\texttt{threading} module can be used in the scripts which use the Python
wrapper around the core implementation, or a C++ driver around the core could
use a higher-level abstraction, like OpenMP as we mentioned earlier, or
explicitly implement threading with pthreads, for example. Achieving this kind
of independence from threading scheme and guaranteeing thread-safety, while not
breaking existing interfaces to the C++ library, was an interesting software
engineering challenge. We solved this by having portions of the API, which were
exposed as thread-safe, maintain their own per-thread state objects. These
state objects are looked up in a C++ Standard Template Library (STL)
\texttt{map}, where thread identification numbers are the keys. The
identification number for a particular thread is found by having that thread
itself query the OS kernel via a system call. This solution does introduce a
small amount of overhead from having a thread inquire about its identification
number on every entry to a function exposed via the API, but it neatly avoids
the problem of breaking existing interfaces, which were written with a single
thread of execution in mind.

\subsection{Data Pump and Parser Operations}

The multi-core machines one encounters in the HPC world may have multiple
memory controllers, where one controller is closer (in terms of signal travel
distance) to one CPU than another CPU. These are \textit{Non-Uniform Memory
Access} (NUMA) architectures. A ramification of working with machines of this
architecture is that memory fetch times may vary significantly depending on
physical address. As bioinformatics software often requires a large memory
footprint to run, it is often found running on these machines. Therefore, if
one is using multiple threads, which may be pinned to various \textit{NUMA
nodes}, the locality of the physical RAM must be taken into consideration. To
this end, we divide our prefetch buffer into a number of segments equal to the
number of threads of execution. Each thread of execution is responsible for
allocating its particular segment of the buffer. The buffer segment is
administered via a state object, maintained on a per-thread basis.

\subsection{Bloom Filter Operations}

The Bloom filter hash tables consume the majority of main memory (see Figure 1) and therefore cannot usefully be split into separate copies among
threads. Rather, a single set of tables must be shared by all of the threads.
This implies that there will be contention among the threads for these
resources. Memory barriers \cite{web:membar} or some form of locking are
needed to prevent two or more threads from attempting to access the same memory
location at the same time. We use atomic addition operations to increment the
counters in the hash tables. These atomic operations \cite{web:atomics} are
supported on a number of platforms by several compiler suites, the GNU
compilers among those, and are not beholden to any particular threading scheme
or library. They establish memory barriers around the operands which they are
to update, thus adding thread-safety to a particular operation.

A performance bottleneck, which we did not address, is the time to write the
hash tables out to storage after k-mer counting is complete. We did not feel
that this was such a high priority because the write-out time is constant for a
given Bloom filter size and is not dependent upon the amount of input data.
For a particular 5 GB data set, which we used for benchmarking, we saw that
k-mer counting took over six times as long as hash table write-out. For even
larger data sets, the ratio becomes more pronounced. That said, we are
ultimately interested in improving performance here too. One possibility is to
amortize the cost of the write-out over the duration of the k-mer counting
phase of operation. The URL-shortener site, bit.ly, has a counting Bloom filter
implementation, called \texttt{dablooms} \cite{web:dablooms}, which achieves
this by memory-mapping its output file to the hash table memory. Adopting their
idea, in conjunction with batch updates of the hash tables, would effectively
give us asynchronous output in bursts over a process' lifetime and chop off the
entire write-out time from the end of execution. Our output is not simply
tables of counts, however, but also includes a header with some metadata;
implementing memory-mapping in light of this fact is an endeavor that needs to
be approached thoughtfully and carefully.

\subsection{Scaling}

Was making the khmer software scalable worth our effort? Yes. Of course, we did
not achieve perfectly linear speedup. But, for every doubling of the number of
cores, we presently get about a factor of 1.9 speedup.

\begin{figure}[ht!]
\centering
\includegraphics[scale=0.75]{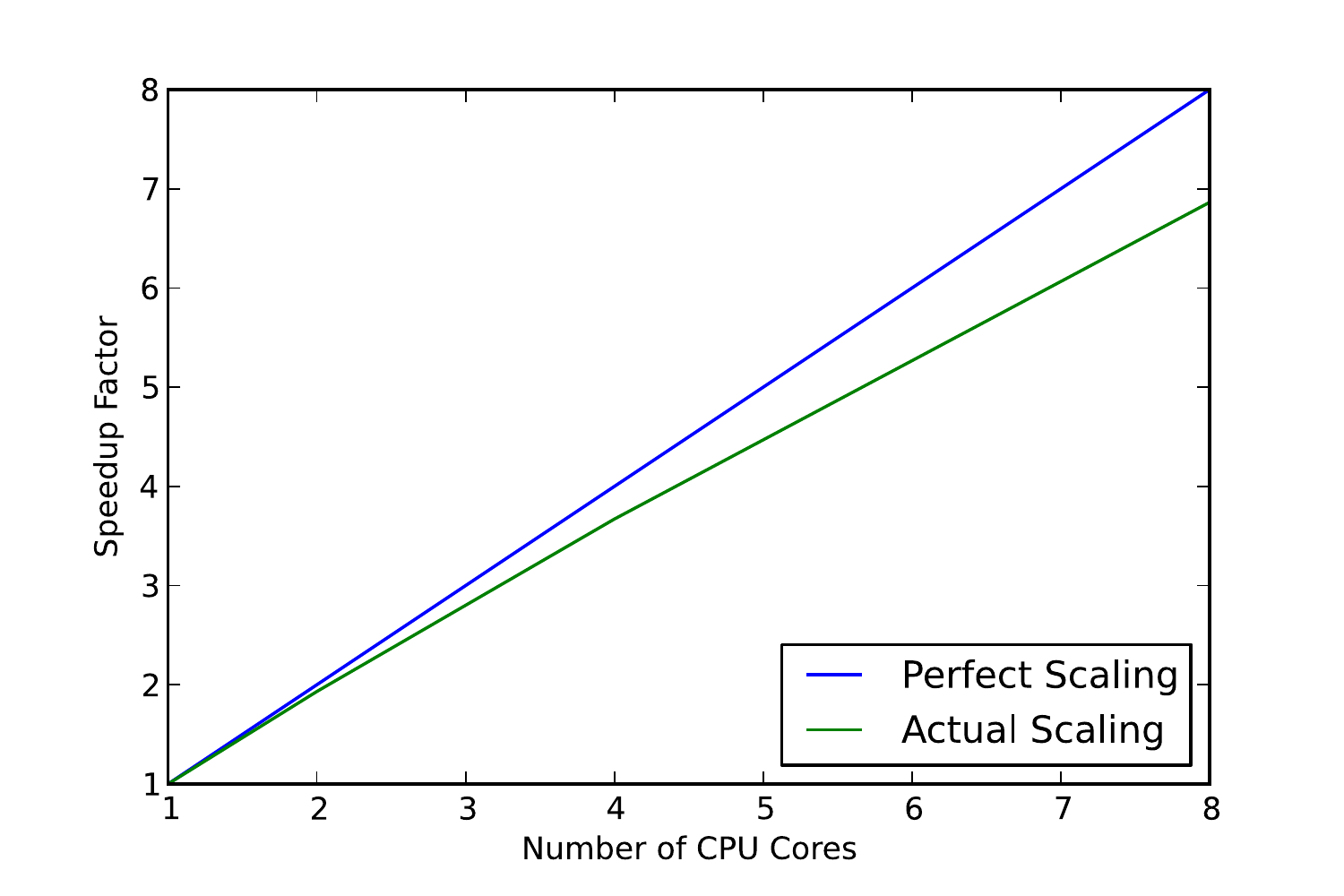}
\caption{Speedup Factor from 1 to 8 CPU Cores}
\label{khmerScaling}
\end{figure}

In parallel computing, one must be mindful of Amdahl's Law \cite{web:Amdahl}
and the Law of Diminishing Returns. The common formulation of Amdahl's Law, in
the context of parallel computing, is $S(N) = \frac{1}{(1 - P) + \frac{P}{N}}$,
where $S$ is the speedup achieved given $N$ CPU cores and, $P$, the proportion
of the code which is parallelized.  For $\lim_{N\to\infty} S = \frac{1}{(1 -
P)}$, a constant.  The I/O bandwidth of the storage system, which the software
utilizes, is finite and non-scalable; this contributes to a non-zero $(1 - P)$.
Moreover, contention for shared resources in the parallelized portion means
that $\frac{P}{N}$ is, in reality, $\frac{P}{N^l}$, where $l < 1$ versus the
ideal case of $l = 1$.  Therefore, returns will diminish over a finite number
of cores even more rapidly.

Using faster storage systems, such as solid-state drives (SSDs) as opposed to
hard-disk drives (HDDs), increases I/O bandwidth (and thus reduces $(1 - P)$),
but that is beyond the purview of software. While we cannot do much about
hardware, we can still try to improve $l$. We think that we can further improve
our access patterns around shared resources, such as the hash table memory, and
that we can better streamline the use of per-thread state objects. Doing these
two things will likely grant us an improvement to $l$.

\section{Conclusion}

The khmer software is a moving target. New features are being added to it
regularly, and we are working on incorporating it into various software stacks
in use by the bioinformatics community. Like many pieces of software in
academia, it started life as an exploratory programming exercise and evolved
into research code.  Correctness was and is a primary goal of the project.
While performance and scalability cannot truly be regarded as afterthoughts,
they yield precedence to correctness and utility. That said, our efforts
regarding scalability and performance have produced good results, including
speedups in single-threaded execution and the ability to significantly reduce
total execution time by employing multiple threads. Thinking about performance
and scalability issues led to the redesign of the data pump and parser
components. Going forward, these components should be able to benefit not only
from scalability but improved maintainability and extensibility.

\section{Future Directions}

Looking forward, once we have addressed the basic performance issues,
we are primarily interested in growing the programmer's API, providing
well tested use cases and documentation, and providing well-characterized
components for integration into larger pipelines. More broadly, we would like
to take advantage of advances in the theory of low-memory data structures to
simplify certain use cases, and we are also interested in investigating
distributed algorithms for some of the more intractable data set challenges
facing us in the near future.

Some additional concerns facing khmer development include an expansion
of the hashing options to allow the use of different hash functions
for single-stranded DNA and the addition of a rolling hash function to
permit k $>$ 32.

We look forward to continuing the development of this software and
hope to have an impact on the Big Data problem facing molecular
biologists and bioinformaticians. We hope that you enjoyed reading
about some high performance, open source software being employed in
the sciences.

\section{Acknowledgements}

We thank Alexis Black-Pyrkosz and Rosangela Canino-Koning for comments and
discussion.

\bibliographystyle{plain}
\bibliography{references}

\begin{thebibliography}{10}

\bibitem{web:dablooms}
bit.ly~software developers.
\newblock dablooms: a scalable, counting {Bloom} filter.
\newblock \url{http://github.com/bitly/dablooms}.

\bibitem{diginorm}
CT~Brown, A~Howe, Q~Zhang, A~Pyrkosz, and TH~Brom.
\newblock A reference-free algorithm for computational normalization of shotgun
  sequencing data.
\newblock In review at PLoS One, July 2012; Preprint at
  http://arxiv.org/abs/1203.4802, 2012.

\bibitem{web:TAU}
A.~D.~Malony et~al.
\newblock {TAU}: {Tuning} and {Analysis Utilities}.
\newblock \url{http://www.cs.uoregon.edu/Research/tau/home.php}.

\bibitem{web:khmer}
C.~Titus~Brown et~al.
\newblock khmer: genomic data filtering and partitioning software.
\newblock \url{http://github.com/ged-lab/khmer}.

\bibitem{web:Valgrind}
Julian~Seward et~al.
\newblock {Valgrind}.
\newblock \url{http://valgrind.org/}.

\bibitem{web:OpenMP}
OpenMP members.
\newblock {OpenMP}.
\newblock \url{http://openmp.org}.

\bibitem{kmer-percolation}
J~Pell, A~Hintze, R~Canino-Koning, A~Howe, JM~Tiedje, and CT~Brown.
\newblock Scaling metagenome sequence assembly with probabilistic de bruijn
  graphs.
\newblock Accepted at PNAS, July 2012; Preprint at
  http://arxiv.org/abs/1112.4193, 2012.

\bibitem{web:Amdahl}
[Various].
\newblock {Amdahl's Law}.
\newblock
  \url{http://en.wikipedia.org/w/index.php?title=Amdahl%27s_law&oldid=51592992%
9}.

\bibitem{web:atomics}
[Various].
\newblock atomic operations.
\newblock
  \url{http://en.wikipedia.org/w/index.php?title=Linearizability&oldid=5116505%
67}.

\bibitem{web:bigdata}
[Various].
\newblock big data.
\newblock
  \url{http://en.wikipedia.org/w/index.php?title=Big_data&oldid=521018481}.

\bibitem{web:BloomFilter}
[Various].
\newblock {Bloom} filter.
\newblock
  \url{http://en.wikipedia.org/w/index.php?title=Bloom_filter&oldid=520253067}.

\bibitem{web:membar}
[Various].
\newblock memory barrier.
\newblock
  \url{http://en.wikipedia.org/w/index.php?title=Memory_barrier&oldid=51764217%
6}.

\bibitem{web:PGO}
[Various].
\newblock profile-guided optimization.
\newblock
  \url{http://en.wikipedia.org/w/index.php?title=Profile-guided_optimization&o%
ldid=509056192}.

\end{thebibliography}
\end{document}